# Waves on Reissner's membrane: a mechanism for the propagation of otoacoustic emissions from the cochlea

Tobias Reichenbach[1], Aleksandra Stefanovic[1], Fumiaki Nin, and A. J. Hudspeth*

Howard Hughes Medical Institute and Laboratory of Sensory Neuroscience, The Rockefeller University, New York, NY 10065-6399, USA

[1]These authors contributed equally to this work.

*To whom correspondence should be addressed
 E-mail: hudspaj@rockefeller.edu
 Phone: +1-212-327-7351
 Fax: +1-212-327-7352

**Summary**

**Sound is detected and converted into electrical signals within the ear. The cochlea not only acts as a passive detector of sound, however, but can also produce tones itself. These otoacoustic emissions are a striking manifestation of the cochlea's mechanical active process. A controversy remains of how these mechanical signals propagate back to the middle ear, from which they are emitted as sound. Here we combine theoretical and experimental studies to show that mechanical signals can be transmitted by waves on Reissner's membrane, an elastic structure within the cochea. We develop a theory for wave propagation on Reissner's membrane and its role in otoacoustic emissions. Employing a scanning laser interferometer, we measure traveling waves on Reissner's membrane in the gerbil, guinea pig, and chinchilla. The results accord with the theory and thus support a role for Reissner's membrane in otoacoustic emissions.**

# Introduction

A healthy ear emits sound that can be recorded by a microphone in the ear canal. In the absence of external sound stimulation such a microphone detects so-called spontaneous otoacoustic emissions (SOAEs), signals at various frequencies that are characteristic of a particular ear and have been proposed for biometric identification (Swabey et al., 2004). An otoacoustic emission can also be evoked by external sound. In response to a pure tone, the ear emits a signal at the same frequency that is termed a stimulus-frequency otoacoustic emission (SFOAE; Kemp, 1978; Robinette and Glattke, 2007; Bergevin et al., 2008). When stimulated with two pure sounds at nearby primary frequencies $f_1$ and $f_2$, the ear produces distortion-product otoacoustic emissions (DPOAEs) at linear combinations of the primary frequencies. Among these the frequencies $2f_1 - f_2$ and $2f_2 - f_1$ are especially prominent (Martin et al., 1998; Robinette and Glattke, 2007; Bergevin et al., 2008). Because of the cochlea's complex mechanics, both the origin of otoacoustic emissions and their mechanism of propagation from the cochlea remain controversial.

The mammalian cochlea acts like an inverse piano to spatially separate frequencies (Pickles, 1996; Ulfendahl, 1997; Robles and Ruggero, 2001). Sound produces an oscillating pressure difference across the basilar membrane inside the cochlea and thus evokes a traveling wave of basilar-membrane displacement. Because the mechanical properties of the basilar membrane change along the cochlea, every point exhibits a resonant frequency that decreases from base to apex. The basilar-membrane wave elicited by a pure tone travels apically until it nears its resonant position, before which it peaks and then declines sharply. The waves elicited by high-frequency sounds peak near the cochlear base and those spawned by low-frequency sounds more apically. This mechanism of frequency selectivity is termed critical-layer absorption because a wave slows upon approaching its resonant position such that it dissipates most of its energy there (Lighthill, 1981).

Signal detection and frequency separation in the cochlea are greatly improved through an active process that provides tuned mechanical amplification of weak signals (Pickles, 1996; Ulfendahl, 1997; Robles and Ruggero, 2001). Specialized outer hair cells sense basilar-membrane vibration and amplify it. The effect of amplification is most pronounced near the peak of the traveling wave, where the basilar-membrane displacements in response to varying sound-pressure levels exhibit a strong compressive nonlinearity. This characteristically nonlinear response indicates operation near an instability such as a Hopf bifurcation (Strogatz, 1994; Wiggins, 1990; Eguíluz et al., 2000; Camalet et al., 2000; Hudspeth et al., 2010). Loss of the active process, for example in a dead cochlea, greatly reduces the peak amplitude and entirely linearizes the response.

Otoacoustic emissions are a hallmark of the active process that disappear when that process is deficient, so they are employed as a clinical test for healthy hearing in newborns (Robinette and Glattke, 2007). Because distortion arises from the nonlinearity owing to cochlear amplification, distortion-product otoacoustic emissions arise near the peaks of the traveling waves elicited by the primary stimulus frequencies $f_1$ and $f_2$ (Robles et al., 1991, 1997; Cooper and Rhode, 1997; Cooper, 1998; Olson, 2004; Dong and Olson, 2005). It remains controversial, however, how a distortion product generated within the cochlea propagates backward to the base (Nobili et al., 2003; Ren, 2004; Shera et al., 2004; Hea et al., 2007; Dong and Olson, 2008; He et al., 2008; Meenderink and van der Heijden, 2010; Sisto, 2011). An understanding of retrograde propagation is complicated by the finding that a distortion-product otoacoustic emission contains two components that differ in their behavior when the primary frequencies $f_1$ and $f_2$ are changed while the ratio $f_2/f_1$ is kept constant (Kemp, 1986, 1999; Knight and Kemp, 2000, 2001; Bergevin et al., 2008). As the primary frequencies are raised, the phase of one component of the distortion-product otoacoustic emission remains approximately constant, whereas the phase of the other component exhibits an increase relative to those of the primary frequencies.

It has been suggested that the two components of a distortion-product otoacoustic emission are generated by distinct mechanisms. Two propositions have been advanced to explain the uniform phase component. First, the generation of distortion by the cochlear nonlinearity probably elicits both forward- and backward-propagating waves on the basilar membrane (de Boer et al., 1986; Kanis and de Boer, 1997; Shera and Guinan, 1999). Waves on the basilar membrane evoked by a pure tone exhibit an approximate scale invariance, executing two to three cycles between the stapes and their peaks regardless of the frequency and direction of travel. As a consequence, a distortion-product otoacoustic emission mediated by a backward-propagating wave exhibits a constant phase that is independent of the primary frequencies. Distortion might alternatively elicit in the cochlear fluid a fast compression wave that transmits a signal (Ren, 2004; He, 2008, 2010). Because the wavelength of such a wave considerably exceeds the length of the cochlea, such a wave would also contribute to the uniform-phase component of an otoacoustic emission.

Only a single mechanism has been proposed to underlie the phase-varying component. The anterograde traveling wave on the basilar membrane produced by cochlear distortion might be reflected near its resonant position and then travel basally (Zweig and Shera, 1995; Shera and Guinan, 1999; Kalluri and Shera, 2001; Talmadge and Dhar, 1999). Reflection is thought to arise from inhomogeneities in the basilar membrane that act as scatterers.

Here we provide an alternative explanation for the emergence of distortion-product otoacoustic emissions. We show that the two components can be explained by waves of two types in the cochlea, one that propagates on the basilar membrane and another that travels on Reissner's membrane. Although both components are produced by nonlinear distortion on the basilar membrane, they propagate in different ways from their generation sites back to the middle ear.

# Results

*Theoretical basis of waves on Reissner's membrane*

Reissner's membrane and the basilar membrane delimit three fluid-filled chambers within the cochlea: scala tympani, scala media, and scala vestibuli (Figure 1A). The mechanosensitive hair cells reside in the organ of Corti on the basilar membrane, which forms one boundary of the scala media. Deflection of the basilar membrane shears the hair bundles of hair cells, which opens mechanically sensitive ion channels and produces electrical responses in these cells. Two specializations of the scala media enhance mechanotransduction by hair cells. First, the scala media contains endolymph, a $K^+$-rich solution that fosters a large cation current through the hair bundles' mechanotransduction channels. Second, the scala media maintains an endocochlear potential of about 80 mV that provides a strong driving force for cations through the mechanotransduction channels.

Although both the basilar membrane and Reissner's membrane separate the specialized endolymph from the perilymph, only the basilar membrane is known to carry traveling waves. As described in the Introduction, anatomical specializations of the basilar membrane—including radial fibers that impose a high stiffness, a width that increases from base to apex, and variation in the size of the organ of Corti—produce traveling waves that peak at frequency-dependent positions. Reissner's membrane, in contrast, lacks such specializations, exhibits a comparatively low impedance, and has therefore been assumed to comply with basilar-membrane motion (Fuhrmann et al., 1987).

Waves might propagate on Reissner's membrane as well. Although the mechanical properties of Reissner's membrane have rarely been studied, Békésy's classical measurements demonstrated a static impedance of Reissner's membrane comparable to that of the basilar membrane near the cochlear apex (Békésy, 1960). The mechanics of the approximately isotropic Reissner's membrane is dominated by surface tension, so waves could occur on it by a mechanism analogous to capillary waves on a water surface.

Consider Reissner's membrane in a coordinate system in which $x$ is the coordinate along the cochlea and $y$ is the radial coordinate across the membrane. The coordinate $z$ then lies perpendicular to $x$ and $z$ such that the membrane is located at $z = 0$ (Figure 1B). Denote by $p_1$ the pressure above and by $p_2$ the pressure below the membrane. A local pressure difference across Reissner's membrane evokes a curvature in its vertical displacement $D_{RM}(x,y)$, which for small deflections satisfies

$$(p_2 - p_1)\big|_{z=0} = -T(\partial_x^2 + \partial_y^2)D_{RM} \tag{1}$$

with the membrane's surface tension $T$ (Landau and Lifshitz, 2007). We consider longitudinal waves in the $x$ direction, for which the membrane exhibits a parabolic shape in the $y$ direction (Figure 1B). For such motion the bending in the $y$ direction makes a contribution of $-T\partial_y^2 D_{RM} = 8TD_{RM}/w^2\big|_{y=0}$, in which $w$ denotes the membrane's width and $y = 0$ its midline. We can then characterize Reissner's membrane by its midline deflection, $D_{RM}\big|_{y=0}$:

$$(p_2 - p_1)\big|_{y=z=0} = -T\left(\partial_x^2 - \frac{8}{w^2}\right)D_{RM}\big|_{y=0}. \tag{2}$$

Stimulation of the membrane at a frequency $f$, and hence an angular frequency $\omega = 2\pi f$, yields a traveling wave analogous to the capillary waves on a water surface owing to surface tension (Lighthill, 1996; Landau and Lifshitz, 2007, Extended Experimental Procedures):

$$D_{RM}\big|_{y=0} = \tilde{D}_{RM} e^{i\omega t - ikx} + c.c., \tag{3}$$

in which $\tilde{D}_{RM}$ is the Fourier component and c.c. denotes the complex conjugate. The wavelength $\lambda$ follows from the wave vector $k$ as $\lambda = 2\pi/k$. In the case of a wavelength less than the height $h$ of each channel, the pressure associated with this wave decays exponentially in the transverse direction. Because the length scale of the exponential decay is provided by the wavelength $\lambda$ (Figure 1C), the presence of the basilar membrane as well as the finite height of the scala vestibuli can be ignored for small wavelengths. The wave vector $k$ then satisfies the dispersion relation

$$2\rho\omega^2 = Tk\left(k^2 + \frac{8}{w^2}\right) \tag{4}$$

in which $\rho$ is the density of the aqueous media.

The width of Reissner's membrane is comparable to the height of the scalae, around 700 μm in rodents, so a wavelength that is smaller than the height is also less than the membrane's width. The parenthetical term in the dispersion relation is therefore dominated by $k^2$ and the relation can be approximated as $2\rho\omega^2 = Tk^3$, from which the wavelength follows as

$$\lambda = 2\pi\left(\frac{T}{2\rho}\right)^{\frac{1}{3}}\omega^{-\frac{2}{3}}. \tag{5}$$

In particular we obtain the scaling $\lambda \sim f^{-2/3}$ for the wavelength's dependence on frequency.

*Measurement of waves on Reissner's membrane*

To test these ideas, we used a scanning laser interferometer to record the midline motion of Reissner's membrane near the cochlear apex from *in vitro* and *in vivo* preparations. Sound stimulation at a single frequency evoked a sinusoidal displacement whose phase $\phi$ varied by multiple cycles over the measured distance of about 1.5 mm. This behavior implies the propagation of a traveling wave (Figure 2A and Supplemental Movie S1). The wavelength $\lambda$ follows as the inverse of the phase slope, $\lambda = (d\phi/dx)^{-1}$, in which the phase is measured in cycles. The phase slope and hence the wavelength varies with frequency: higher frequencies lead to steeper phase changes and hence smaller wavelengths (Figure 2A,B).

We measured waves on Reissner's membrane in three rodents: the gerbil, guinea pig, and chinchilla. The wavelengths for a given frequency were comparable across species (Figure 2B). Moreover, the frequency dependences of the wavelength within each species confirmed the scaling $\lambda \sim f^{-2/3}$ for frequencies above 1 kHz. The measured wavelengths allowed us to infer the surface tension of Reissner's membrane, which is about 120 mN·m$^{-1}$ for the gerbil, 180 mN·m$^{-1}$ for the guinea pig, and 270 mN·m$^{-1}$ for the chinchilla. These values are of the same order of magnitude as previous measurements of the surface tension of Reissner's membrane (Békésy, 1960; Steele, 1974).

We also quantified the amplitude of waves on Reissner's membrane elicited by sound stimulation (Figure 2C). The sensitivity, defined as the wave's displacement amplitude normalized by the sound pressure applied in the ear canal, was about 10 nm·Pa$^{-1}$ for frequencies below 5 kHz. This value is comparable to the sensitivity of waves on the basilar membrane in the absence of the active process or at high sound-pressure levels (Robles and Ruggero, 2001). We conclude that, in a passive cochlea, sound stimulation elicits a wave on the Reissner's membrane at a comparable amplitude to the wave on the basilar membrane.

*Modes of propagation on the fluid-coupled basilar and Reissner's membranes*

Even for high-frequency stimulation, a wave on the basilar membrane has a wavelength comparable to or greater than the height of the scalae (Ulfendahl, 1997; Robles and Ruggero, 2001). Such a wave is therefore influenced both by Reissner's membrane and by the boundaries at the walls of the scalae. At frequencies below 1 kHz a wave on Reissner's membrane also has a wavelength exceeding the height of the scalae (Figure 2B), so such a wave interacts with the basilar membrane and with the upper and lower cochlear walls. We next consider the consequences of these interactions.

Consider a two-dimensional model of the cochlea in which $x$ is the coordinate along the cochlear length and $z$ the coordinate normal to the membranes (Figure 3A, Extended Experimental Procedures). The hydrodynamics follows from Laplace equations for the pressures in the scala vestibuli, scala media, and scala tympani, which we denote by respectively $p_1$, $p_2$, and $p_3$:

$$\Delta p_1 = 0, \ \Delta p_2 = 0, \ \Delta p_3 = 0. \tag{6}$$

We have approximated the fluid as incompressible and the flow as laminar. Boundary conditions for the Laplace equations arise at the upper and lower walls of the cochlea, where the transverse fluid velocities must vanish:

$$\partial_z p_1 \big|_{z=3h} = \partial_z p_3 \big|_{z=0} = 0. \tag{7}$$

Additional boundary conditions arise at Reissner's membrane and the basilar membrane. The pressure difference across Reissner's membrane evokes a velocity $V_{RM}$ there and the pressure difference across the basilar membrane produces a velocity $V_{BM}$. We consider a wave propagating at angular frequency $\omega$ with a local wave vector $k$. The specific acoustic impedances $Z_{RM}(\omega,k) = -iT(k^2 + 8/w^2)/\omega$ of Reissner's membrane (Equation 2) and $Z_{BM}(\omega)$ of the basilar membrane then relate the pressure differences to the membrane velocities:

$$(p_2 - p_1)|_{z=2h} = Z_{RM} V_{RM},$$
$$(p_3 - p_2)|_{z=h} = Z_{BM} V_{BM}. \quad (8)$$

The imaginary component of the basilar membrane's impedance varies spatially. The membrane's stiffness decreases from the cochlear base to the apex, whereas the organ of Corti and the tectorial membrane grow in size, conferring an increasing mass. The wavelength and amplitude of a wave thus vary spatially:

$$V_{RM}(x) = \tilde{V}_{RM}(x) \exp\left[i\omega t - i\int_0^x dx' k(x')\right] + c.c.,$$
$$V_{BM}(x) = \tilde{V}_{BM}(x) \exp\left[i\omega t - i\int_0^x dx' k(x')\right] + c.c.. \quad (9)$$

These equations describe a wave traveling on both membranes with a local wave vector $k(x)$ and complex local amplitudes $\tilde{V}_{RM}(x)$ and $\tilde{V}_{BM}(x)$. Analysis of Equations 6 together with the boundary conditions, Equations 7 and 8, shows that the local wave vector $k(x)$ obeys the dispersion relation

$$\left\{\frac{ik(x)Z_{RM}}{\rho\omega}\sinh[k(x)h] - 2\cosh[k(x)h]\right\}\left\{\frac{ik(x)Z_{BM}(x)}{\rho\omega}\sinh[k(x)h] - 2\cosh[k(x)h]\right\} = 1. \quad (10)$$

Details of this analysis are relegated to the Extended Experimental Procedures.

An important property of this dispersion relation is its invariance under a change of sign for $k(x)$. A particular solution $k(x)$ of the dispersion relation thus implies that $-k(x)$ is a solution as well: for each forward-traveling wave there exists an analogous backward-traveling wave and *vice versa*.

Each solution $k(x)$ to the dispersion relation, Equation 10, defines a wave that propagates both on Reissner's membrane and on the basilar membrane and hence represents a mode of motion of the coupled membranes. The ratio of the Reissner's membrane motion to that of the basilar membrane is given by

$$\frac{\tilde{V}_{RM}(x)}{\tilde{V}_{BM}(x)} = \frac{ik(x)Z_{BM}(x)}{\rho\omega}\sinh[k(x)h] - 2\cosh[k(x)h]. \tag{11}$$

Numerical analysis of Equation 10 reveals two fundamental solutions $k_a(x)$ and $k_b(x)$ that reflect the two fundamental degrees of freedom in the cochlea, namely the motions of the two membranes. In the basal region of the cochlea, and for frequencies above 1 kHz, the two modes adopt simple forms. First, and as shown in the previous section, Reissner's membrane then sustains a wave whose wavelength is smaller than the height of the scalae and that accordingly does not penetrate significantly into the membrane's surrounding fluids. This wave operates in the short-wavelength limit $|k_a(x)|h \gg 1$. Approximating $\sinh[k_a(x)h] \approx \cosh[k_a(x)h] \gg 1$ in the dispersion relation, Equation 10, we obtain the solution $k_a(x) = \pm 2i\rho\omega/Z_{RM}$ in agreement with Equation 4 and 5. It follows from Equation 11 that the basilar-membrane motion evoked by this wave is negligible. Because the propagation of this wave is, to good approximation, determined by the impedance of Reissner's membrane alone, we refer to this mode as the Reissner's membrane mode (Figure 3B).

A second, long-wavelength mode $k_b(x)$ exists whose wavelength exceeds the height of the channels, $|k_b(x)|h \ll 1$. In this instance we can approximate $\sinh[k_b(x)h] \approx k_b(x)h$ and $\cosh[k_b(x)h] \approx 1$. Because the basilar-membrane impedance near the base considerably exceeds that of Reissner's membrane, $Z_{BM}(x) \gg Z_{RM}$, we find that $k_b(x) = \pm\sqrt{-3i\rho\omega/[2hZ_{BM}(x)]}$. The motion of Reissner's membrane approximately equals that of the basilar membrane, which reflects the long wavelength of this mode as well as the high compliance of Reissner's membrane relative to that of the basilar membrane. Because the propagation of this mode reflects predominantly the impedance of the basilar membrane, we refer to this mode as the basilar-membrane mode (Figure 3C).

Because the impedance of Reissner's membrane shows little or no spatial variation, the amplitude of a wave on that structure remains essentially constant along the cochlea. A wave using the basilar-membrane mode, however, changes in amplitude as the impedance of the basilar membrane varies. The change in amplitude can be computed from the energy flow associated with this wave: for a passive system the vibration amplitudes of Reissner's membrane and the basilar membrane must change in such a way that the energy flow at each longitudinal location remains constant (Steele and Taber, 1979; Lighthill, 1981). In conjunction with Equation 11, this condition defines the vibration amplitudes of the two membranes and can be solved numerically. An analytical approximation is feasible because the basilar membrane bears long waves and because its impedance significantly exceeds that of Reissner's membrane. As a result, the amplitude of the basilar-membrane motion changes in proportion to $\sqrt{k_b(x)}$ and the amplitude of motion by Reissner's membrane follows from Equation 11 (Figure 2C and Extended Experimental Procedures). Because the vibration of Reissner's membrane is comparable to that of the basilar membrane, measurements from Reissner's membrane can be employed to characterize the basilar-membrane mode (Rhode, 1987; Hao and Khanna, 1999).

The above arguments reveal that, near the base of the cochlea, Reissner's membrane has little effect on the basilar-membrane mode. Insofar as motion of the basilar membrane is concerned, Reissner's membrane may therefore be neglected, as has indeed been done in most previous cochlear models. Near the cochlear apex, however, this assumption fails for two reasons. First, when its wavelength exceeds the height of the scalae, a wave traveling on Reissner's membrane interacts with the basilar membrane. Analytical as well as numerical solutions reveal that the wavelength then scales as $\lambda \sim f^{-1}$ (Figure 2B and Extended Experimental Procedures). Second, the impedance of the basilar membrane near the apex is comparable to that of Reissner's membrane (Békésy, 1960). At low frequencies and near the apex, both modes are therefore influenced by the impedances of Reissner's membrane as well as of the basilar membrane. This situation, which we shall not discuss further, confounds an interpretation of these modes as purely a basilar-membrane mode and a Reissner's membrane mode. In particular, the influence of Reissner's membrane may pose a problem for a mechanical resonance of the basilar membrane near the apex and suggests the presence of an alternative tuning mechanism there (Reichenbach and Hudspeth, 2010a,b).

*Distortion products*

Distortion products are produced by a nonlinear response of the basilar membrane. A pure tone evokes a wave that travels apically toward a resonant position near which it peaks and then decays sharply. Near the resonant position the membrane's response becomes strongly nonlinear:

$$\left(p_3 - p_2\right)\big|_{z=h} = Z_{BM} V_{BM} + A V_{BM}^3, \tag{12}$$

in which we have assumed a cubic nonlinearity supplementing the linear response and in which $A$ is a proportionality coefficient. When stimulated at two frequencies $f_1$ and $f_2$, a cubic nonlinearity produces distortion frequencies such as $2f_1 - f_2$ and $2f_2 - f_1$ (Extended Experimental Procedures). The basilar membrane is excited at those distortion frequencies at positions near the peaks of the waves of the primary frequencies $f_1$ and $f_2$.

Which waves are elicited by local stimulation of the basilar membrane from within the cochlea? Consider a force at a distortion frequency $\omega$ that acts on the basilar membrane at a single position $x_0$. Such a force is proportional to $\cos(\omega t)\delta(x - x_0)$, in which $\delta(x - x_0)$ represents a Dirac delta function that is centered at $x = x_0$ and vanishes elsewhere. Employing techniques developed in elementary-particle physics, we have found an analytical solution for the pressures in the different scalae that follow from this type of forcing (Figure S1 and Extended Experimental Procedures). The resulting pressures $p_1^{(G;x_0,\omega)}$, $p_2^{(G;x_0,\omega)}$, and $p_3^{(G;x_0,\omega)}$ are known as Green's functions and are commonly employed to solve inhomogeneous differential equations. In our case, they satisfy the Laplace Equations 6 as well as the boundary conditions, Equations 7 and 8, with the boundary condition at the basilar membrane adjusted to

$$\left(p_3^{(G;x_0,\omega)} - p_2^{(G;x_0,\omega)}\right)\big|_{z=h} = Z_{BM} V_{BM} + p_F \cos(\omega t)\delta(x - x_0) \tag{13}$$

to reflect forcing of the basilar membrane at a pressure amplitude $p_F$. The pressures $p_1^{(G;x_0,\omega)}$, $p_2^{(G;x_0,\omega)}$, and $p_3^{(G;x_0,\omega)}$ in response to forcing at position $x_0$ represent two waves. First, forcing of the basilar membrane unsurprisingly elicits a wave on that structure. Second, and less intuitively, a force on the basilar membrane also evokes a wave on Reissner's membrane. How does this wave arise? As found in the previous discussion, the basilar-membrane mode has a large wavelength and thus travels both on Reissner's membrane and on the basilar membrane. To evoke the basilar-membrane mode alone would require a force to act on both membranes in a specific proportion. A force originating only on the basilar membrane inevitably excites a second wave on Reissner's membrane.

To examine the effect of the two modes on distortion-product otoacoustic emissions, we have used a cochlear model with realistic parameter values to compute the pressure amplitude evoked at the stapes through forcing of the basilar membrane at various positions $x_0$ (Figure 4A,B). There is an important difference between the responses that result from the two modes. The pressure amplitude at the stapes that is induced by the basilar-membrane mode decays sharply when the position of forcing lies apical to the place of the characteristic frequency. This drop occurs because of critical-layer absorption on the basilar membrane: a wave of any particular frequency cannot propagate on that structure apical to its resonant position, nor can forcing there elicit such a wave. No such complication arises with disturbances propagating by the Reissner's membrane mode, which can advance both basally and apically from their site of generation (Figure 4B).

We have also computed the amplitudes and phases of the two modes created by distortion when the cochlea is stimulated at two nearby frequencies $f_1$ and $f_2$. The nonlinearity in the basilar membrane's response produces distortion not just at a single position, but over the extended cochlear segment where the nonlinear response dominates the linear one (Equation 12). The resulting pressures $p_1$, $p_2$, and $p_3$ are accordingly a superposition of the pressures $p_1^{(G;x_0,\omega)}$, $p_2^{(G;x_0,\omega)}$, and $p_3^{(G;x_0,\omega)}$ emerging from forcing at each position $x_0$ and at different frequencies $\omega$:

$$p_n = \int_0^\infty d\omega \int_0^1 dx_0 \, p_n^{(G;x_0,\omega)} p_F^{-1} A \overline{V_{BM}^3}(x_0,\omega) \quad \text{for } n = 1, 2, 3 \tag{14}$$

in which $\overline{V_{BM}^3}(x_0,\omega)$ is the Fourier component of $V_{BM}^3(x_0,t)$ at angular frequency $\omega$. Because the velocity $V_{BM}$ of the basilar membrane depends on the pressures, $\rho\partial_t V_{BM} = -\partial_y p_2 = -\partial_y p_3$, Equation 14 cannot be solved directly. Recordings of otoacoustic emissions show, however, that the sound-pressure levels for the distortion products lie well below those of the primary frequencies (Kemp, 1978; Martin et al., 1998; Knight and Kemp, 2001; Bergevin et al., 2008). Because the pressures from distortion products represent small perturbations, we can approximate the pressures that appear on the right-hand side of Equation 14 by the pressures that result from stimulation at the primary frequencies $f_1$ and $f_2$. This type of approximation, which was introduced into wave theory by Max Born in the context of quantum mechanics, represents the first contribution in a perturbation series for the solution of Equation 14 (Sakurai, 1994).

Our computations confirm that distortion products originate primarily within a narrow region of the cochlea (Figure 4C). For the lower sideband frequency $2f_1 - f_2$, waves propagating by both modes emerge predominantly from the region where the basilar-membrane waves at the primary frequencies peak and overlap. The same holds for a wave moving by the Reissner's membrane mode at the upper sideband frequency $2f_2 - f_1$. However, the basilar-membrane mode at $2f_2 - f_1$ behaves differently. Because the basilar-membrane waves elicited by the primary frequencies peak apically to the characteristic place for the frequency $2f_2 - f_1$, a basilar-membrane wave at that frequency cannot propagate there. The $2f_2 - f_1$ emission thus arises more basally, near its characteristic place. Because this region lies basally to the peak regions of the primary frequencies, the basilar-membrane wave at the upper sideband frequency $2f_2 - f_1$ is excited less and thus has a smaller amplitude than that at the lower sideband frequency $2f_1 - f_2$. In fact, the amplitude of the basilar-membrane mode at $2f_2 - f_1$ is even smaller than that of the Reissner's membrane mode at that frequency.

For both the upper and the lower sidebands, we have computed the total amplitude of the two waves and their dependence on the ratio $f_2/f_1$ of the primary frequencies (Figure 5A). The lower-sideband emission is dominated by the basilar-membrane mode whereas the upper sideband is dominated by the Reissner's membrane mode. This difference results primarily from a change in the amplitude of the basilar-membrane mode. As explained above, the basilar-membrane mode for an upper-sideband emission does not arise within the peak region of the primaries but more basally and thus has a reduced amplitude. The amplitude of the Reissner's membrane mode is similar for the lower- and upper-sideband frequencies but declines as the ratio of the primary frequencies increases because the basilar-membrane waves induced by the primary frequencies then overlap less. A previous experimental study of the amplitude of both components and their dependence on the ratio $f_2/f_1$ indeed obtained very similar results (Figure 5 in Knight and Kemp, 2001).

As the primary frequencies $f_1$ and $f_2$ change at a constant ratio $f_2/f_1$, the phase behavior of the distortion-product emission through the Reissner's membrane mode differs dramatically from that through the basilar-membrane mode (Figure 5B). The emission through the basilar-membrane mode maintains an almost constant phase. The approximate scale invariance for frequencies above 1 kHz indeed implies that, independently of the frequency of stimulation, the basilar-membrane wave elicited by a pure tone travels two to three cycles to reach its resonant position. A basilar-membrane wave produced by the cochlear nonlinearity thus travels a similar number of cycles basally from its site of generation until it reaches the stapes, again independently of the frequency. No such argument applies to the Reissner's membrane mode. As the primary frequencies and hence the distortion-product frequency increase, the waves on Reissner's membrane decrease in wavelength (Figure 2B and Equation 5). The waves therefore undergo a larger number of cycles and thus acquire a progressively greater phase delay while traveling from their generation site to the stapes. Although this effect is slightly reduced because the generation site of the distortion product shifts basally for higher frequencies, a phase lag of several cycles nonetheless accumulates as the primary frequencies change by a few kilohertz.

### *Measurement of distortion products on Reissner's membrane*

By using a scanning laser interferometer to record from the apical cochlear turns of living chinchillas, we have measured the propagation of distortion products on Reissner's membrane.

Stimulation at two frequencies $f_1$ and $f_2$ above 1 kHz results in a signal at the cubic distortion frequency $2f_2 - f_1$ (Figure 6A,D). Because the characteristic frequency of auditory-nerve fibers in the cochlear region at which we recorded is below 1 kHz (Eldredge, 1981), these distortion products are created basally to our site of measurement. We therefore expect to observe a forward-traveling wave in the Reissner's membrane mode. Signals in the basilar-membrane mode should not reach the measurement site, for both the upper- and lower-sideband distortion products occur at frequences of at least 1 kHz.

Scanning along the midline of the membrane demonstrates a progressive phase decrease that signals a forward-traveling wave (Figure 6C,F). The wavelength given by the inverse of the phase slope is smaller for a higher distortion-product frequency and agrees with our single-frequency measurements of waves on Reissner's membrane (Figure 2). These interferometric measurements therefore confirm that the basilar membrane's nonlinear response evokes a traveling wave in the Reissner's membrane mode.

## Discussion

Our results show that otoacoustic emissions can emerge from the cochlea in two distinct ways that correspond to two modes of propagation on the parallel, fluid-coupled Reissner's membrane and basilar membrane. For emissions from the basal portion of the cochlea at frequencies above 1 kHz, the two modes have intuitive interpretations. The basilar-membrane mode is determined predominantly by the basilar membrane's impedance and involves almost equal displacements of both membranes. The Reissner's membrane mode travels almost exclusively on Reissner's membrane with a negligible displacement of the basilar membrane. Although the active force from cochlear outer hair cells acts directly on the basilar membrane but not on Reissner's membrane, we have shown that it excites both the basilar-membrane and the Reissner's membrane modes.

Although traveling waves in the basilar-membrane mode have been extensively measured and analyzed (Lighthill, 1981; Ulfendahl, 1997; Robles and Ruggero, 2001), the present study is the first to describe waves in the Reissner's membrane mode. We have measured the waves on Reissner's membrane in different rodent species and found agreement of the inferred dispersion relation with our theoretical prediction. We have also shown that these waves can be produced by distortion on the basilar membrane.

We have demonstrated that the two components of a distortion-product otoacoustic emission—which emerge through the two wave modes in the cochlea—differ in their phase behavior when the primary frequencies are changed at a constant ratio. The phase of the emission through the basilar-membrane mode remains approximately constant, whereas that involving the Reissner's membrane mode changes by multiple cycles as the primary frequencies are swept across a few octaves. Previous experiments have indeed measured two such components (Kemp, 1986, 1999; Knight and Kemp, 2000, 2001; Bergevin et al., 2008). We therefore identify the constant-phase component with the emission that propagates in the basilar-membrane mode and the phase-varying component with the emission that travels in the Reissner's membrane mode.

Our theory allows us to quantify the amplitude of the two components in a distortion-product otoacoustic emission. We confirm that the lower-sideband emission, $2f_1 - f_2$, is dominated by the constant-phase component whereas the upper-sideband signal, $2f_2 - f_1$, is carried predominantly by the phase-varying component. Experiments have previously revealed this remarkable behavior (Kemp, 1986, 1999; Knight and Kemp, 2000, 2001; Bergevin et al., 2008). In particular, a detailed study of the amplitude of both components and their dependence on the ratio $f_2/f_1$ obtained results very similar to ours (Figure 5 in Knight and Kemp, 2001). Although for frequency ratios close to one the amplitude of the phase-varying component does not change much between the upper- and lower-sideband emissions, the amplitude of the constant-phase component is significantly greater for the lower sideband. This distinct behavior emerges naturally in our theory because a basilar-membrane wave cannot travel across its resonant position whereas a wave on Reissner's membrane can propagate along the whole extent of the cochlea.

We have also quantified the generation sites of the distortion products. Both components of a lower-sideband emission, as well as the phase-varying component of an upper-sideband emission, originate in the region where the traveling waves associated with the primary frequencies peak and overlap. The constant-phase component of the upper-sideband emission, however, arises more basally, near the characteristic place for the frequency $2f_2 - f_1$. This difference in generation sites accords with experimental measurements (Martin et al., 1998).

The emission of a distortion product through a backward-traveling wave on the basilar membrane has been challenged by some recent experiments but is supported by others (Ren, 2004; He, 2008, 2010; Dong and Olson, 2008; Meenderink and van der Heijden, 2010). Our results show that a distortion traveling backward through the basilar-membrane mode displays characteristic behaviors, both regarding the strength of the resulting emission and its phase, that are consistent with experimental observations of the uniform-phase component (Fig. 5). Distortion might alternatively elicit a fast pressure wave if the cochlear active process were to produce a local volume change, for example in outer hair cells (Wilson, 1980). Future experiments should clarify whether the active process can yield such a volume change or whether distortion excites a backward-traveling basilar-membrane mode.

In this study we have focused for three reasons on frequencies above 1 kHz. First, because the electronic noise in microphones increases at low frequencies, most otoacoustic emissions have been measured at frequencies exceeding 1 kHz. Second, the mechanics of the basilar membrane has been studied predominantly in the basal region; the mechanics of the cochlear apex appears to differ (Cooper and Rhode, 1995; Khanna and Hao, 1999, 2000; Zinn et al., 2000; Robles and Ruggero, 2001; Temchin et al., 2008; Reichenbach and Hudspeth, 2010a,b). Third, and in agreement with the previous point, we have shown here that cochlear waves at frequencies below 1 kHz are influenced by the properties of both Reissner's membrane and the basilar membrane, which confounds a simple interpretation of the modes. Different cochlear mechanics near the apex and near the base may underlie the experimental differences in otoacoustic emissions at low and high frequencies (Knight and Kemp, 2001; Shera and Guinan, 1999). This issue is a promising subject for future investigations.

Although we have focused on the distortion-product otoacoustic emissions that have been studied most extensively, our theory should hold for other types of otoacoustic emissions as well. We expect that future experiments will delineate two components in stimulus-frequency and spontaneous otoacoustic emissions.

Otoacoustic emissions serve as an important clinical measure for hearing in newborns (Robinette and Glattke, 2007). Because our study offers a better understanding of the mechanisms of otoacoustic emissions, we hope that it will allow more refined conclusions from such tests about the normal functioning or impairment of hearing.

## Experimental Procedures

### *Cochlear preparations*

Measurements of waves on Reissner's membrane were performed on cochlear preparations both *in vivo* and *in vitro*. For an *in vitro* experiment we euthanized a guinea pig (*Cavia porcellus*) 6-8 weeks of age or a Mongolian jird or gerbil (*Meriones unguiculatus*) 5-8 weeks of age with sodium pentobarbital (Nembutal, Lundbeck Inc., Deerfield, IL) and dissected the cochlea together with the middle ear. The bulla was glued to a plastic support and opened to afford optical access to the cochlear apex. A piece of cochlear bone 0.5-1.5 mm in length was removed from the apex to expose the underlying Reissner's membrane.

For an *in vivo* measurement we used standard preparative techniques (Cooper and Rhode, 1997, 1997; Ren, 2002) on a guinea pig 6-8 weeks of age or a chinchilla (*Chinchilla lanigera*) 8 weeks of age. As in the *in vitro* experiments we gained access to Reissner's membrane through a fenestra in the apical turn of the cochlea.

*Stimulation*

Waves were initiated in three different ways. For some of the *in vitro* guinea pig preparations we made an opening into the scala media of the second cochlear turn. We advanced a micropipette through this fenestra in parallel with the basilar membrane until it contacted Reissner's membrane. Using a piezoelectric stack (P-883.11, Physik Instrumente, Karlsruhe, Germany), we then stimulated Reissner's membrane directly by imposing a sinuosidal oscillation on the micropipette.

In the remaining experiments on gerbils *in vitro* and in all of the single-frequency experiments *in vivo*, we delivered sound signals with a loudspeaker (ES1, Tucker-Davis Technologies, Alachua, FL) that was connected to the external ear canal through a tube.

For distortion-product measurements we separately generated two primary frequencies that were delivered through independent loudspeakers (ES1, Tucker-Davis Technologies) connected to the ear canal through a branched tube.

*Sound calibration*

We employed for calibration a sensitive microphone (4939, Brüel & Kjær, Nærum, Denmark) with a defined ratio of output voltage to sound-pressure level. The microphone was inserted into a coupler that was connected by independent tubes to an animal's ear canal and to a miniature loudspeaker. We then stimulated the speaker with different voltages and recorded the ensuing sound-pressure levels.

*Laser interferometry*

We measured the vibrations of Reissner's membrane along its midline with a scanning laser interferometer (OFV-501, Polytec, Waldbronn, Germany). To increase the membrane's reflectivity, we placed on it glass beads 10 μm in diameter.

*Data collection and analysis*

Stimulation and recording were performed with two synchronized audio signal processing boards (RX6, Tucker-Davis Technologies) and LabVIEW 7.0 (National Instruments, Austin, TX) operating at digital output and sampling intervals of 10 μs. Data analysis was conducted with Mathematica 6.0 (Wolfram Research, Champaign, IL).

## Acknowledgments

We thank B. Fabella and J. A. N. Fisher for assistance with laser interferometry and the members of our research group for comments on the manuscript. This research was supported by grant DC000241 from the National Institutes of Health. T. R. holds a Career Award at the Scientific Interface from the Burroughs Wellcome Fund; A. J. H. is an Investigator of Howard Hughes Medical Institute.

# Figure Legends

**Figure 1. Waves on Reissner's membrane.** (A) Reissner's membrane (RM) and the basilar membrane (BM) delineate three fluid-filled chambers—the scala vestibuli (SV), scala media (SM), and scala tympani (ST)—within the cochlear duct. The scala media contains $K^+$-rich endolymph that baths the hair cells of the organ of Corti (OC) and the overlying tectorial membrane (TM). (B) A schematic diagram depicts a wave of wavelength $\lambda$ on Reissner's membrane, positioned in the $x, y, z$ coordinate system used in our theoretical calculations. (C) In a wave on Reissner's membrane, fluid particles move in circular trajectories (blue) when the wavelength $\lambda$ is smaller than the height of the scalae. The radius of these trajectories decays exponentially with the distance from the membrane with a space constant proportional to the wavelength.

**Figure 2. Measurements of waves on Reissner's membrane.** (A) Sound stimulation of an *in vivo* preparation of the guinea pig's cochlea vibrates the Reissner's membrane as measured near the cochlear apex. The phase accumulation over the region of measurement indicates the presence of traveling waves propagating from base to apex (left to right). (B) Waves on the Reissner's membranes of different rodents display a similar dependence of wavelength on the stimulus frequency. For stimulation at frequencies exceeding 1 kHz the wavelength decreases as $f^{-2/3}$. The black line, which represents the behavior expected from theory, reveals a crossover from this scaling at high frequencies to scaling as $f^{-1}$ at low frequencies. This transition occurs near a wavelength $\lambda = 2h$ or a frequency of 1 kHz. The measurements from chinchillas and those marked (1) from guinea pigs were performed *in vivo*; the experiments on gerbils and those marked (2) from guinea pigs employed *in vitro* preparations. (C) The sensitivity of Reissner's membrane waves to acoustic stimulation is about 10 nm·Pa$^{-1}$ for frequencies up to 5 kHz and declines for greater frequencies. Four different experiments, represented by different symbols,

were performed on guinea pig cochleas *in vivo*. For an animation of the waves measured on Reissner's membrane see Movie S1.

**Figure 3. Two modes of propagation on Reissner's membrane and the basilar membrane.** (A) In a schematic diagram of a two-dimensional cochlear model, acoustic stimulation displaces the stapes at the oval window (bold arrow); the round window (thin arrow) moves subsequently in response to the propagating pressure wave. (B) A wave in the Reissner's membrane mode propagates without variation in amplitude or wavelength and does not evoke a significant displacement of the basilar membrane. (C) In contrast, a disturbance moving in the basilar-membrane mode propagates on both membranes. As the wave approaches its resonant position, the vibration amplitudes of both membranes increase whereas the wavelength and speed decrease. The amplitudes decay sharply beyond the peaks. The displacement of Reissner's membrane is comparable to that of the basilar membrane basal to the peak but then declines as the fluid coupling between the membranes falls with decreasing wavelength.

**Figure 4. Cochlear origin of distortion-product otoacoustic emissions.** The panels depict the distortion-product otoacoustic emissions computed to emerge from stimulation at 60 dB SPL at frequencies $f_1$ and $f_2$ (solid lines) or $\hat{f}_1$ and $\hat{f}_2$ (dashed lines). These frequencies are arranged such that the same distortion product, $f = 2000$ Hz, emerges either as the lower sideband $2f_1 - f_2$ or as the upper sideband $2\hat{f}_1 - \hat{f}_2$. The ratios of the primary frequencies in the two instances are $f_2/f_1 = 1.3$ and $\hat{f}_2/\hat{f}_1 = 1.6$. (A) The amplitudes of basilar-membrane waves for each of the stimulus frequencies are shown along with the amplitude of the wave that would emerge for acoustic stimulation at frequency $f$. The sites of maximal overlap of the waves elicited by stimuli at $f_1$ and $f_2$, as well as the corresponding loci for $\hat{f}_1$ and $\hat{f}_2$, are indicated in this and the two following panels (dotted black lines). (B) Driving the basilar membrane at a frequency $f$ and at varying positions $x_0$ evokes retrograde traveling waves in the Reissner's membrane mode (green) and basilar-membrane mode (red). The pressures at the stapes are shown relative to the pressure

$p_F$ at the site of stimulation. The Reissner's membrane mode can be excited from any cochlear position, whereas the basilar-membrane mode is active only basal to the resonant position. (C) Simultaneous stimulation with sound at frequencies $f_1$ and $f_2$ elicits pressures at the stapes at the distortion frequency $f$ from similar extended cochlear regions (solid lines) for emissions through the two modes. Simultaneous stimulation at $\hat{f}_1$ and $\hat{f}_2$ produces distortion responses (dashed lines) through the two modes that differ in their relative amplitudes and cochlear origins owing to the inability of the distortion products to propagate on the basilar membrane apically to their characteristic places. See Figure S1 and Table S1 for additional details.

**Figure 5. Experimentally observable results from a computational model.** (A) The pressures of the distortion products at the stapes differ strikingly for emissions through the two modes. When the primary frequencies are near one another, emission through the Reissner's membrane mode (green) has an approximately equal amplitude for the upper and the lower sidebands. Emission through the basilar-membrane mode (red), however, is much stronger at the lower sideband than at the upper sideband. (B) The emissions at the frequency $2f_1 - f_2$ through the two modes show distinct phase changes as the primary frequencies vary at a constant ratio. The phase of the emission through the basilar-membrane mode remains approximately constant, whereas that of the emission through the Reissner's membrane mode changes by several cycles.

**Figure 6. Measurement of distortion-product propagation along Reissner's membrane.** We show exemplary results from one of three successful *in vivo* measurements from the chinchilla. (A) The frequency spectrum during stimulation at the primary frequencies $f_1 = 1.3$ kHz and $f_2 = 1.6$ kHz shows the lower-sideband cubic distortion product $2f_1 - f_2 = 1$ kHz. The upper-sideband cubic distortion is weak and comparable to the noise floor. (B) The distortion product disappears after the animal has been sacrificed. (C) Scanning along Reissner's membrane at the distortion-product frequency $2f_1 - f_2$ reveals a progressive decrease of the signal's phase, an indication of a traveling wave moving from base to apex. (D) Stimulation at $f_1 = 2.5$ kHz and $f_2 = 3$ kHz evokes the lower-sideband cubic distortion product $2f_1 - f_2 = 2$ kHz. (E) The distortion product vanishes in a dead animal. (F) The phase decline again implies that the distortion product $2f_1 - f_2$ propagates as a forward traveling wave on Reissner's membrane.

## Supplemental Information

The Supplemental Information include one Supplemental Figure, one Supplemental Table, one Supplemental Movie, and Extended Experimental Procedures.

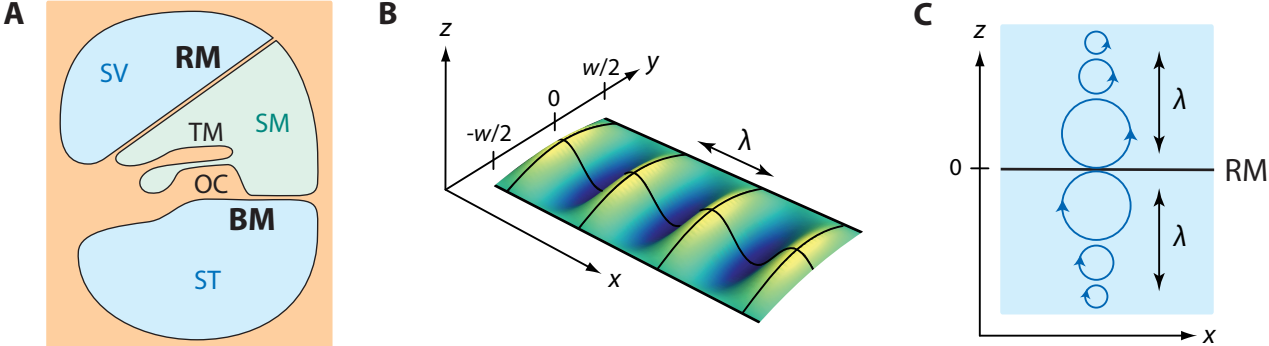

Reichenbach, Stefanovic, Nin, and Hudspeth

Figure 2

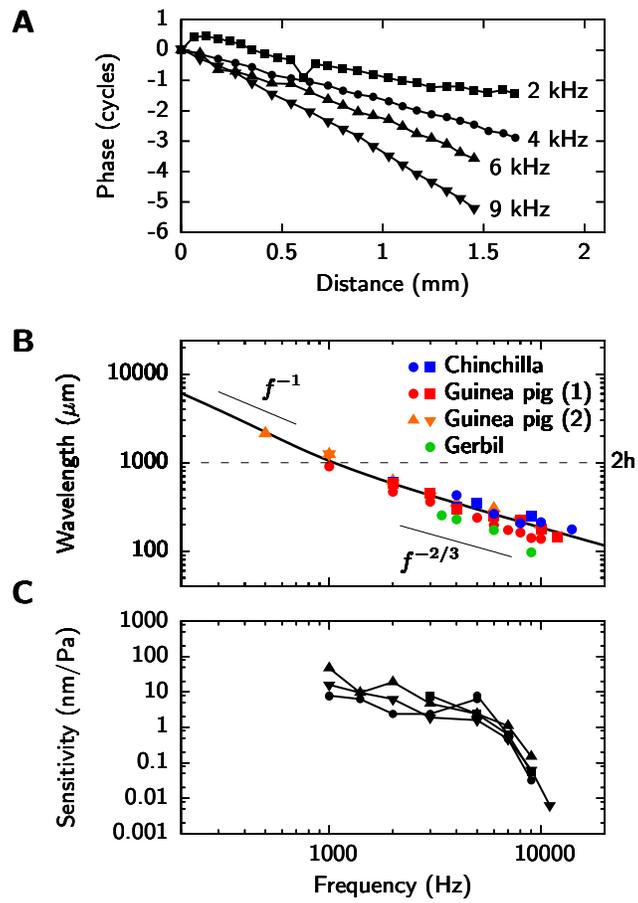

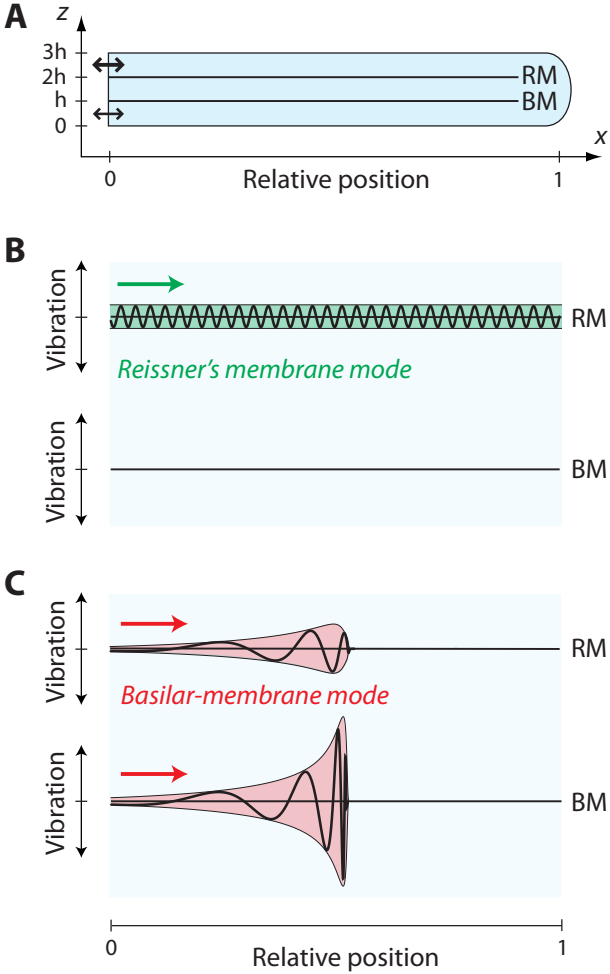

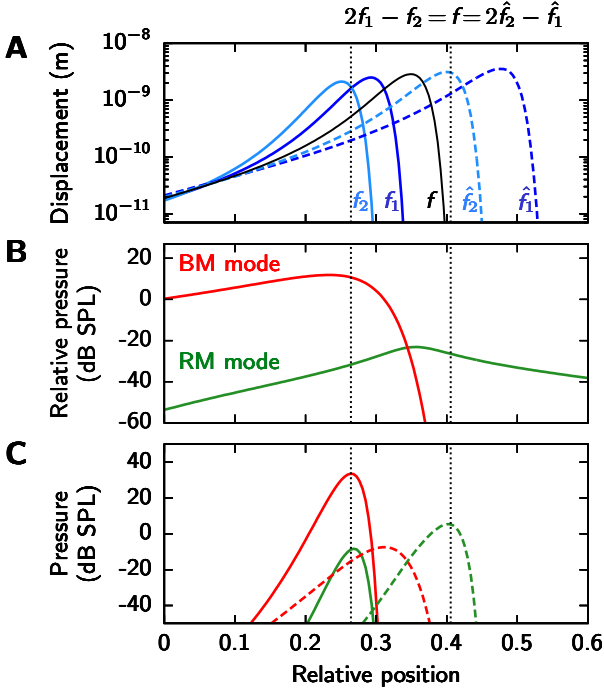

Reichenbach, Stefanovic, Nin, and Hudspeth

Figure 5

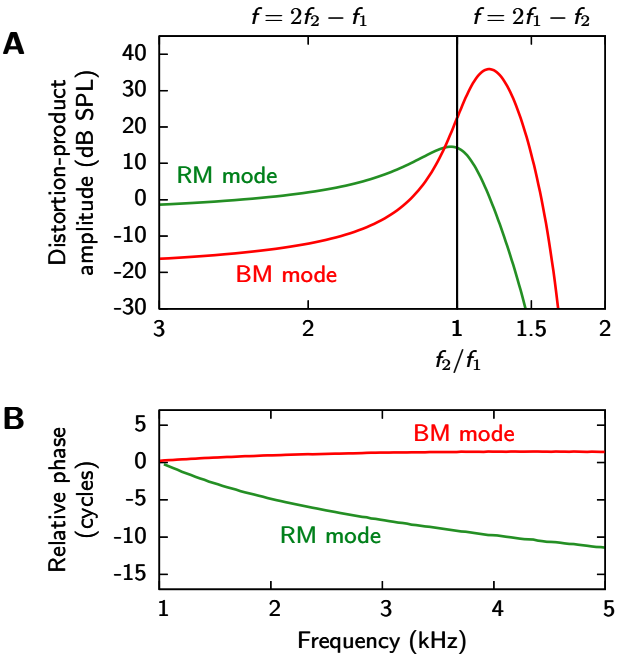

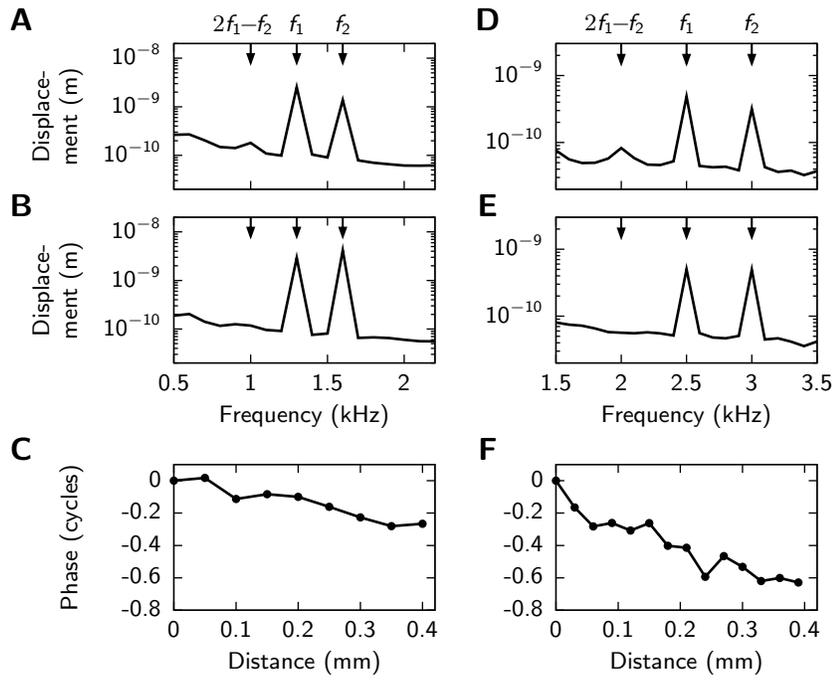

# Supplemental Information

# Waves on Reissner's membrane: a mechanism for the propagation of otoacoustic emissions from the cochlea

Tobias Reichenbach[1], Aleksandra Stefanovic[1], Fumiaki Nin, and A. J. Hudspeth*

Howard Hughes Medical Institute and Laboratory of Sensory Neuroscience,
The Rockefeller University, New York, NY 10065-6399, USA

[1]*These authors contributed equally to this work.*

*To whom correspondence should be addressed
E-mail: hudspaj@rockefeller.edu
Phone: +1-212-327-735
Fax: +1-212-327-73521

# 1. Extended Experimental Procedures

## A. Fluid dynamics of the waves on Reissner's membrane

We consider Reissner's membrane within an $x,y,z$ coordinate system (Figure 1B) and describe the hydrodynamics in the plane $y = 0$ that lies perpendicular to the membrane and along its midline. For small displacements and incompressible fluids, the pressures $p_1$ and $p_2$, respectively above and below the membrane, satisfy the Laplace equations

$$\Delta p_1 = 0, \ \Delta p_2 = 0. \tag{S1}$$

We consider a high angular stimulation frequency $\omega$ for which the height of the scalae exceeds the wavelength. The pressures must therefore vanish far from the boundaries, which is fulfilled in the ansatz

$$\begin{aligned} p_1 &= -\tilde{p} e^{i\omega t - ikx - kz} + c.c., \\ p_2 &= \tilde{p} e^{i\omega t - ikx + kz} + c.c., \end{aligned} \tag{S2}$$

in which $c.c.$ represents the complex conjugate. The pressures satisfy the Laplace relations (Equations S1) and decay exponentially with distance from the membrane with a length scale proportional to the wavelength $\lambda = 2\pi/k$.

Reissner's membrane imposes a boundary condition (Equation 2). Because $\rho \partial_t^2 X_{\text{RM}} = -\partial_z p_1 \big|_{z=0} = -\partial_z p_2 \big|_{z=0}$, we obtain the dispersion relation (Equation 4).

## B. Wave propagation on the parallel Reissner's membrane and basilar membrane

To solve the Laplace relations (Equation 6) together with the boundary conditions (Equations 7 and 8), we employ the ansatz

$$\begin{aligned} p_1 &= \tilde{p}_1(x)\cosh\left[\partial_x b(x)(z - 3h)\right] e^{i\omega t - i\omega b(x) - \partial_x b(x)h} + c.c., \\ p_2 &= \left\{\tilde{p}_2^u(x)\cosh\left[\partial_x b(x)(z - 2h)\right] + \tilde{p}_2^d(x)\cosh\left[\partial_x b(x)(z - h)\right]\right\} e^{i\omega t - i\omega b(x) - \partial_x b(x)h} + c.c., \\ p_3 &= \tilde{p}_3(x)\cosh\left[\partial_x b(x)z\right] e^{i\omega t - i\omega b(x) - \partial_x b(x)h} + c.c.. \end{aligned} \tag{S3}$$

Because the local wave vector $k(x)$ is related to the phase $b(x)$ by $k(x) = \omega \partial_x b(x)$, the phase may be expressed through the wave vector, $b(x) = \int_0^x dx' k(x')/\omega$.

The pressures yield the velocities of Reissner's membrane and the basilar membrane,

$$-\rho \partial_t V_{RM} = \partial_z p_1\big|_{z=2h} = \partial_z p_2\big|_{z=2h},$$
$$-\rho \partial_t V_{BM} = \partial_z p_2\big|_{z=h} = \partial_z p_3\big|_{z=h}. \tag{S4}$$

Applying the WKB approximation, we consider an expansion in powers of the angular frequency $\omega$ (Steele & Taber, 1979; Lighthill, 1996; Reichenbach & Hudspeth, 2010b). A high frequency implies a small wavelength $\lambda(x)$, a length scale over which the basilar-membrane impedance $Z_{BM}(x)$ varies little. The spatial variation of the pressure amplitudes, $\partial_x p_n$ ($n = 1,2,3$), then results predominantly from the derivative of the phase, $\partial_x b(x)$: the corresponding terms are of order $\omega$ whereas terms that involve $\partial_x \tilde{p}_1$, $\partial_x \tilde{p}_2^u$, $\partial_x \tilde{p}_2^d$, $\partial_x \tilde{p}_3$ and $\partial_x^2 b(x)$ are of the smaller order 1. To leading order $\omega^2$ we hence find $\partial_x^2 p_n = -[\partial_x b(x)]^2 p_n$ ($n = 1,2,3$). Because $\partial_z^2 p_n = [\partial_x b(x)]^2 p_n$, the pressures satisfy the Laplace relations (Equations 6).

To leading order $\omega^2$ the boundary conditions yield

$$\tilde{p}_2^d(x) = -\tilde{p}_1,$$
$$\tilde{p}_2^u(x) = -\tilde{p}_3, \tag{S5}$$

as well as

$$\frac{\tilde{p}_3}{\tilde{p}_1} = \frac{ik(x)Z_{RM}}{\rho\omega}\sinh[k(x)h] - 2\cosh[k(x)h],$$
$$\frac{\tilde{p}_1}{\tilde{p}_3} = \frac{ik(x)Z_{BM}(x)}{\rho\omega}\sinh[k(x)h] - 2\cosh[k(x)h]. \tag{S6}$$

The last two equations for the ratio of the pressure amplitudes $\tilde{p}_1$ and $\tilde{p}_3$ must agree, which gives the dispersion relation (Equation 10).

Because the basilar-membrane mode exhibits a large wavelength, $|k(x)|h \ll 1$, we can approximate $\sinh[k(x)h] \approx k(x)h$ and $\cosh[k(x)h] \approx 1$ to obtain

$$k_{1/2}^2(x) = \frac{\rho\omega}{hZ_{RM}Z_{BM}(x)}\left\{-i[Z_{RM} + Z_{BM}(x)] \pm i\sqrt{Z_{RM}^2 - Z_{RM}Z_{BM}(x) + Z_{BM}^2(x)}\right\}. \tag{S7}$$

When the impedance of Reissner's membrane is much below that of the basilar membrane, $Z_{RM} \ll Z_{BM}(x)$, we find that

$$k^2(x) = -\frac{3i\rho\omega}{2hZ_{BM}(x)} \tag{S8}$$

and the wave vector $k(x)$ depends on only the basilar membrane's impedance. Because the wavelength of this mode is much greater than the height of the channels, the system can be regarded as one-dimensional and the WKB approximation yields an amplitude of basilar-membrane vibration that varies in proportion to $\sqrt{k(x)}$ (Steele and Taber, 1979; Lighthill, 1996; Reichenbach and Hudspeth, 2010b).

The impedances of Reissner's membrane and the basilar membrane are comparable near the cochlear apex. Because stimulation at frequencies below 1 kHz elicits large wavelengths for both modes, the respective wave vectors follow from Equation S7. The impedance of Reissner's membrane is dominated by the membrane's transverse flexion, $Z_{RM} \approx -8iT/(\omega w^2)$, and the wavelength of the corresponding wave mode is thus inversely proportional to the frequency, $\lambda \sim f^{-1}$.

## C. Green's functions

The Green's functions, the pressures $p_1^{(G;x_0,\omega)}$, $p_2^{(G;x_0,\omega)}$, and $p_3^{(G;x_0,\omega)}$, fulfill the Laplace relations (Equations 6) together with the boundary conditions (Equations 7 and 8), but the boundary condition at the basilar membrane is given by Equation 13. The WKB approximation again facilitates the solution. As shown above, a wave's local wave vector follows from the local impedance alone, irrespective of putative impedance changes. We therefore start by considering a uniform basilar-membrane impedance $Z_{BM}$ and make the ansatz

$$p_1^{(G;x_0,\omega)} = \int_{-\infty}^{\infty} dk\, G_1(k) \cosh[k(z-3h)] e^{i\omega t - ik(x-x_0) - kh} + c.c.,$$

$$p_2^{(G;x_0,\omega)} = \int_{-\infty}^{\infty} dk \left\{ G_2^u(k) \cosh[k(z-2h)] + G_2^d(k) \cosh[k(z-h)] \right\} e^{i\omega t - ik(x-x_0) - kh} + c.c., \quad (S9)$$

$$p_3^{(G;x_0,\omega)} = \int_{-\infty}^{\infty} dk\, G_3(k) \cosh[kz] e^{i\omega t - ik(x-x_0) - kh} + c.c..$$

From the boundary conditions $\partial_z p_1^{(G;x_0,\omega)}\big|_{z=2h} = \partial_z p_2^{(G;x_0,\omega)}\big|_{z=2h}$ and $\partial_z p_2^{(G;x_0,\omega)}\big|_{z=h} = \partial_z p_3^{(G;x_0,\omega)}\big|_{z=h}$, we obtain $G_2^d(k) = -G_1(k)$ and $G_2^u(k) = -G_3(k)$. Because the Dirac delta function can be represented as

$$\delta(x - x_0) = \frac{1}{2\pi} \int_{-\infty}^{\infty} dk\, e^{-ik(x-x_0)}, \quad (S10)$$

we compute

$$G_1(k) = \frac{p_F e^{kh}}{2\pi L(k)},$$

$$G_3(k) = \left[\frac{ikZ_{RM}}{\rho\omega}\sinh(kh) - 2\cosh(kh)\right] \frac{p_F e^{kh}}{2\pi L(k)}, \quad (S11)$$

in which $L(k)$ is defined as

$$L(k) = \left[\frac{ikZ_{RM}}{\rho\omega}\sinh(kh) - 2\cosh(kh)\right]\left[\frac{ikZ_{BM}}{\rho\omega}\sinh(kh) - 2\cosh(kh)\right] - 1. \quad (S12)$$

With this notation the dispersion relation (Equation 10) reads $L(k) = 0$.

In considering the propagation of distortion products, we are interested in waves far from their generation site $x_0$. The integrals in Equations S9 can then be calculated by closing the contour of integration in the complex plane (Figure S1). Complex analysis informs us that only the poles of the integrand contribute to an integral along such a closed path. Poles occur at those values $k$ for which $L(k)$ vanishes, and hence at the values $\pm k_a$ and $\pm k_b$ with $k_a, k_b > 0$ that describe the two wave modes in the cochlea. If the impedances $Z_{BM}$ and $Z_{RM}$ involve friction, the solutions $\pm k_a$ and $\pm k_b$ possess small imaginary parts and are located in the second and fourth quadrants of the complex plane (Figure S1). For $x < x_0$ we can close the integration contour in the upper half plane, and obtain contributions from $-k_a$ and $-k_b$ that describe retrograde waves. In the opposite case, when $x > x_0$, the contour can be closed in the lower half plane to yield contributions from $k_a$ and $k_b$ and forward-traveling waves.

Because we are interested in the retrograde waves that reach the stapes, we consider $x < x_0$. Denote as $p_n^{(G,a;x_0,\omega)}$ ($n = 1,2,3$) the contribution to the pressure $p_n^{(G;x_0,\omega)}$ from the pole at $-k_a$ and denote as $p_n^{(G,b;x_0,\omega)}$ ($n = 1,2,3$) the contribution from the pole at $-k_b$. The pressures $p_n^{(G,a;x_0,\omega)}$ therefore represent the pressures of the Reissner's membrane mode and the pressures $p_n^{(G,b;x_0,\omega)}$ those of the basilar-membrane mode. We find $p_n^{(G;x_0,\omega)} = p_n^{(G,a;x_0,\omega)} + p_n^{(G,b;x_0,\omega)}$ with

$$p_1^{(G,a;x_0,\omega)} = 2\pi i \left[ \partial_k G_1^{-1}(k) \Big|_{k=-k_a} \right]^{-1} \cosh[k_a(z-3h)] e^{i\omega t + ik_a(x-x_0) + k_a h} + c.c.,$$

$$p_2^{(G,a;x_0,\omega)} = -2\pi i \left\{ \left[ \partial_k G_3^{-1}(k) \Big|_{k=-k_a} \right]^{-1} \cosh[k_a(z-2h)] \right.$$
$$\left. + \left[ \partial_k G_1^{-1}(k) \Big|_{k=-k_a} \right]^{-1} \cosh[k_a(z-h)] \right\} e^{i\omega t + ik_a(x-x_0) + k_a h} + c.c.,$$
(S13)

$$p_3^{(G,a;x_0,\omega)} = 2\pi i \left[ \partial_k G_3^{-1}(k) \Big|_{k=-k_a} \right]^{-1} \cosh[k_a z] e^{i\omega t + ik_a(x-x_0) + k_a h} + c.c.,$$

and the pressures $p_n^{(G,b;x_0,\omega)}$ follow analogously.

In the actual cochlea the basilar-membrane impedance $Z_{BM}(x)$ varies with the longitudinal position $x$. As elaborated above, the local wave vectors $k_a$ and $k_b$ also depend on the position $x$. In the WKB approximation the pressure amplitudes vary as $1/\sqrt{k(x)}$ (Steele and Taber, 1979; Lighthill, 1996; Reichenbach and Hudspeth, 2010b). Because in the WKB approximation, and to leading order, only the local wave vector $k(x)$ contributes to the derivatives of the pressures, one verifies that adjusting the pressures in Equation S13 in proportion to $1/\sqrt{k(x)}$ solves the Laplace relations (Equation 6) with the stated boundary conditions (Equations 7, 8, and 13). For the retrograde waves at $x < x_0$ we find

$$p_1^{(G,a;x_0,\omega)} = 2\pi i \sqrt{\frac{k_a(x_0)}{k_a(x)}} \left[\partial_k G_1^{-1}(k)\Big|_{k=-k_a(x_0)}\right]^{-1} \cosh[k_a(x)(z-3h)] e^{i\omega t + i\int_x^{x_0} dx' k_a(x') + k_a(x)h} + c.c.,$$

$$p_2^{(G,a;x_0,\omega)} = -2\pi i \sqrt{\frac{k_a(x_0)}{k_a(x)}} \left\{\left[\partial_k G_3^{-1}(k)\Big|_{k=-k_a(x_0)}\right]^{-1} \cosh[k_a(x)(z-2h)]\right.$$

$$\left. + \left[\partial_k G_1^{-1}(k)\Big|_{k=-k_a(x_0)}\right]^{-1} \cosh[k_a(x)(z-h)]\right\} e^{i\omega t + i\int_x^{x_0} dx' k_a(x') + k_a(x)h} + c.c.,$$

$$p_3^{(G,a;x_0,\omega)} = 2\pi i \sqrt{\frac{k_a(x_0)}{k_a(x)}} \left[\partial_k G_3^{-1}(k)\Big|_{k=-k_a(x_0)}\right]^{-1} \cosh[k_a(x)z] e^{i\omega t + i\int_x^{x_0} dx' k_a(x') + k_a(x)h} + c.c..$$

(S14)

The pressures $p_n^{(G,b;x_0,\omega)}$ as well as the case $x > x_0$ follow analogously.

### D. Distortion products

The pressure waves produced by nonlinear distortion can be computed through Equation 14 from the Green's functions (Equations S14). This equation contains the Fourier component $\overline{V_{BM}^3}(x_0,\omega)$ from which $V_{BM}^3(x_0,t)$ follows as

$$V_{BM}^3(x_0,t) = \int_0^\infty d\omega \overline{V_{BM}^3}(x_0,\omega) e^{i\omega t} + c.c..$$ (S15)

The Fourier component $\overline{V_{BM}^3}(x_0,\omega)$ can be expressed through the Fourier component $\tilde{V}_{BM}(x_0,\omega)$ of $V_{BM}(x_0,t)$:

$$\overline{V_{BM}^3}(x_0,\omega) = \left[\tilde{V}_{BM}(x_0) * \tilde{V}_{BM}(x_0) * \tilde{V}_{BM}(x_0)\right](\omega)$$ (S16)

in which $*$ denotes the convolution defined by

$$(f*g)(\omega) = \int_{-\infty}^\infty d\omega' f(\omega')g(\omega-\omega') = \int_0^\infty d\omega' \left[f(\omega')g(\omega-\omega') + f^*(\omega')g(\omega+\omega')\right].$$ (S17)

The last equality holds when $f(\omega) = f^*(-\omega)$, as is the case when $f(\omega)$ represents the Fourier component of a real-valued function.

To compute the retrograde waves at the cubic distortion frequencies $2f_1 - f_2$ and $2f_2 - f_1$, we consider stimulation of the cochlea at the two primary frequencies $f_1$ and $f_2$. In the linear, passive cochlea the basilar-membrane response then contains only those two frequencies:

$$\tilde{V}_{BM}(x_0,\omega) = V_{BM}^{(1)}(x_0)\delta(\omega-\omega_1) + V_{BM}^{(2)}(x_0)\delta(\omega-\omega_2).$$ (S18)

Upon inserting Equation S18 into Equation S16 we find that the basilar-membrane inputs at $f_1$ and $f_2$ produce responses at linear combinations, specifically at frequencies $f \in I$ for which the set $I$ is $I = \{f = \pm f_i \pm f_j \pm f_k\}$ with $i,j,k \in \{1,2\}$ and $f > 0$:

$$\overline{V_{\text{BM}}^3}(x_0, \omega) = \sum_{\substack{\omega' = 2\pi f \\ f \in I}} S^{(\omega')}(x_0) \delta(\omega - \omega'). \tag{S19}$$

The amplitudes at the distortion frequencies $2f_1 - f_2$ and $2f_2 - f_1$ are

$$\begin{aligned} S^{(2\omega_1 - \omega_2)}(x_0) &= 3\left[V_{\text{BM}}^{(1)}(x_0)\right]^2 \left[V_{\text{BM}}^{(2)}(x_0)\right]^*, \\ S^{(2\omega_2 - \omega_1)}(x_0) &= 3\left[V_{\text{BM}}^{(2)}(x_0)\right]^2 \left[V_{\text{BM}}^{(1)}(x_0)\right]^*. \end{aligned} \tag{S20}$$

This distortion elicited by the linear, passive basilar-membrane velocity represents the Born approximation to the full, nonlinear Equation 14.

### E. Parameter values

We model a cochlea 35 mm in length with a maximal best frequency of $f_{\text{max}} = 30$ kHz at its base and a minimal best frequency of $f_{\text{min}} = 50$ Hz at its apex. The longitudinal position $x$ is measured in units of the cochlear length such that $x = 0$ denotes the base and $x = 1$ the apex. The maximal and minimal frequencies define an exponential map $f_0(x)$ of best frequencies in the cochlea in which $f_0(x)$ matches $f_{\text{max}}$ at the base and $f_{\text{min}}$ at the apex.

The specific acoustic impedance $Z_{\text{BM}}(x)$ of the basilar membrane follows from the stiffness, viscosity, and mass. We consider a strip of the basilar membrane with a width of 8 μm, the width of one hair cell. This strip has an area of $A_{\text{BM}}(x) = w_{\text{BM}}(x) \cdot 8$ μm, in which $w_{\text{BM}}(x)$ denotes the membrane's width as a function of the longitudinal position $x$. The impedance follows as

$$Z_{\text{BM}} = A_{\text{BM}}^{-1}(x)\left[-iK(x)/\omega + \mu(x) + i\omega m(x)\right], \tag{S21}$$

in which $K(x)$ is the stiffness, $\mu(x)$ the drag coefficient, and $m(x)$ the mass of the basilar-membrane strip.

At each longitudinal position in the cochlea, the mass and stiffness define a resonant frequency $f_{res}(x) = (2\pi)^{-1}\sqrt{K(x)/m(x)}$. We assume that, in the basal region of the cochlea, this resonant frequency equals the best frequency $f_0(x)$ and hence consider a stiffness $K(x)$ proportional to $f_0(x)$ and a mass $m(x)$ inversely proportional to $f_0(x)$. We choose a maximal stiffness $K(x=0) = 1$ N·m$^{-1}$ at the base and a mass according to $f_{res}(x) = f_0(x)$. To represent viscous damping we assume that the drag coefficient $\mu(x)$ is proportional to the membrane's width $w_{BM}(x)$ with a proportionality coefficient of 0.015 N·s·m$^{-2}$.

The nonlinearity that produces distortion results from an active process that counteracts viscous damping. We assume that the active process produces a force that is proportional to the basilar-membrane displacement. The coefficient $A$ in Equation 12 is thus inversely proportional to $\omega^3$ as well as to $A_{BM}(x)$: $A = 5 \cdot 10^{11} \cdot \omega^{-3} \cdot A_{BM}^{-1}$ kg·m$^{-2}$·s$^{-2}$.

The remaining parameter values are summarized in Table S1.

## 2. Supplemental figure titles and legends

**Figure S1. Computation of the Green's functions, Related to Figure 4.** The functions $p_1^{(G;x_0,\omega)}$, $p_2^{(G;x_0,\omega)}$, and $p_3^{(G;x_0,\omega)}$ can be evalulated through integration of Equations S9 in the complex plane. When $x < x_0$ the integration contour can be closed in the upper half plane and yields contributions from $-k_a$ and $-k_b$ that describe retrograde waves. In the opposite case, when $x > x_0$, the contour can be closed in the lower half plane to provide contributions from $k_a$ and $k_b$ and hence forward-traveling waves.

## 3. Supplemental movie titles and legends

**Movie S1. Interferometric recordings of waves propagating on Reissner's membrane, Related to Figure 2.** The images portray the movement along the midline of a segment about 1.5 mm in length near the apex of the guinea pig's cochlea. As quantified in Figure 2, the wavelength decreases with increasing stimulus frequency.

## 4. Supplemental tables

**Table S1. Parameter values used in the numerical computation, Related to Figure 4.**

| Parameter | Description | Value |
| --- | --- | --- |
| $h$ | Height of each of the three scalae | 500 μm |
| $\rho$ | Fluid density | $10^3$ kg·m$^{-3}$ |
| $T$ | Surface tension of Reissner's membrane | 220 mN·m$^{-1}$ |
| $w$ | Width of Reissner's membrane | 700 μm |
| $w_{BM}(x)$ | Width of the basilar membrane | $(100+400 \cdot x)$ μm |

Reichenbach, Stefanovic, and Hudspeth

Figure S1

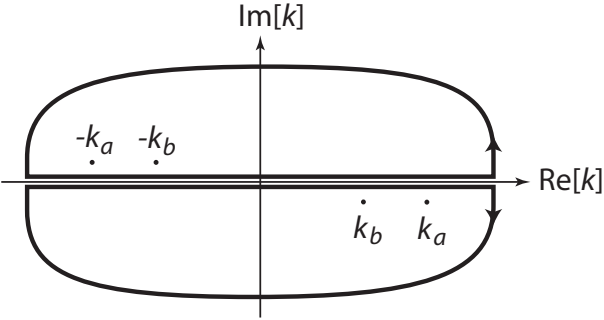